\documentclass[onecolumn]{jaise2e}     

\usepackage[T1]{fontenc}
\usepackage{times}%
\usepackage{amsmath}
\usepackage{graphicx}
\usepackage{dcolumn}

\pubyear{0000}
\volume{0}
\firstpage{1}
\lastpage{1}

\begin{document}
\twocolumn

\begin{frontmatter}

\title{Performance Optimization of Multiple Interconnected Heterogeneous Sensor Networks via Collaborative Information Sharing}
\runningtitle{Perf. Opt. of Mult. Interconnected Heterogeneous Sensor Networks via Collaborative Information Sharing}


\author[A]{\fnms{Sougata} \snm{Pal}},
\author[A]{\fnms{Simon} \snm{Oechsner}}
\author[A]{\fnms{Boris} \snm{Bellalta}\thanks{Corresponding author. E-mail: boris.bellalta@upf.edu}}
\author[A]{\fnms{Miquel} \snm{Oliver}}
\runningauthor{S. Pal, et al.,}
\address[A]{Department of Information and Communication Technologies. Universitat Pompeu Fabra, Barcelona. e-mail:\{name.surname\}@upf.edu}

\maketitle

\begin{abstract}
Interconnecting multiple sensor networks is a relatively new research field which has emerged in the Wireless Sensor Network domain. Wireless Sensor Networks (WSNs) have typically been seen as logically separate, and few works have considered interconnection and interaction between them. Interconnecting multiple heterogeneous sensor networks therefore opens up a new field besides more traditional research on, e.g., routing, self organization, or MAC layer development. Up to now, some approaches have been proposed for interconnecting multiple sensor networks with goals like information sharing or monitoring multiple sensor networks. In this paper, we propose to utilize inter-WSN communication to enable Collaborative Performance Optimization, i.e., our approach aims to optimize the performance of individual WSNs by taking into account measured information from others. The parameters to be optimized are energy consumption on the one hand and sensing quality on the other.
\end{abstract}


\end{frontmatter}


\section{Introduction}

Wireless Sensor Networks (WSNs) \cite{YiMu08} are wireless networks whose primary task is information gathering by monitoring parameters of the environment. They consist of a number of sensor nodes each gathering information and reporting it back to their sink. The observed parameters can take a variety of forms, such as temperature, humidity, pollution, traffic, volcano and earthquake activities, structural deficiencies, bird flight path patterns or even enemy movements on battlefields. Due to this rich and dynamic field of applications WSNs have become an active research field in the area of wireless communications. 

The main research focus for WSNs is on creating smarter, cheaper and more intelligent sensors, as well as more energy efficient networks. But even although there has been tremendous progress, there remain numerous unsolved issues. Moreover, due to their varying application scenarios, individual sensor networks differ from each other, e.g, in terms of their individual protocol stack or network functionality. This is partially due to the fact that different vendors are manufacturing different kinds of sensor network nodes, so every sensor network has its own MAC, routing and transport protocols. Due to this heterogeneity, sensor networks are typically considered as individual entities being logically separated from each other. 

As a result, research on WSNs focuses more on internal issues of single networks, such as routing, energy management or MAC protocol development. Extensive work has already been done in those areas, and development continues. However, the interaction between multiple heterogeneous spatially distributed sensor networks is an issue that has not received much attention. Therefore, sharing information between multiple sensor networks, inter WSN communication, and connecting multiple sensor networks via a feasible architecture have been started to be investigated recently.   

In this work, we describe a novel approach to utilize an inter-WSN communication architecture to optimize the performance of individual WSNs. It can therefore be seen as a complementary work to all performance optimizations focused on single networks without any additional information, e.g., MAC or routing protocols, or cross layer optimizations for the sensor network protocol stack. In our approach, we consider the heterogeneous nature of different types of sensor networks and do not try to generate a global, centrally controlled optimization strategy. We believe it is neither feasible nor practical to propose a common MAC or routing protocol for every sensor network, or to have a common cross layer optimization method for the entire architecture to optimize the performance.

Instead, we propose that individual sensor networks improve their performance by utilizing reports from other sensor networks, taking into account all relevant information to take the best decision locally. To this end, we propose both a global architecture that allows to find and establish contact with these additional information sources, as well as a method how to use this information to reduce energy consumption and to increase the sensing quality of individual networks. To the best of our knowledge, this approach of using reported information from other WSNs to optimize local performance has not been considered before.

The remainder of this paper is organized as follows: Section \ref{sec:benefits} describes the basic idea and model for improving the performance of a WSN using external information. Section \ref{sec:arch} provides the description of our proposed architecture, while Section \ref{sec:usecases} contains realistic use cases for our architecture. Section \ref{sec:relwork} gives the an overview on the related work, and finally Section \ref{sec:conc} summarizes our conclusions and provides an outlook on our future work.


\section{Benefits of Collaboration}\label{sec:benefits}

\begin{figure*}[t]
\centering
\includegraphics[width=0.9\textwidth]{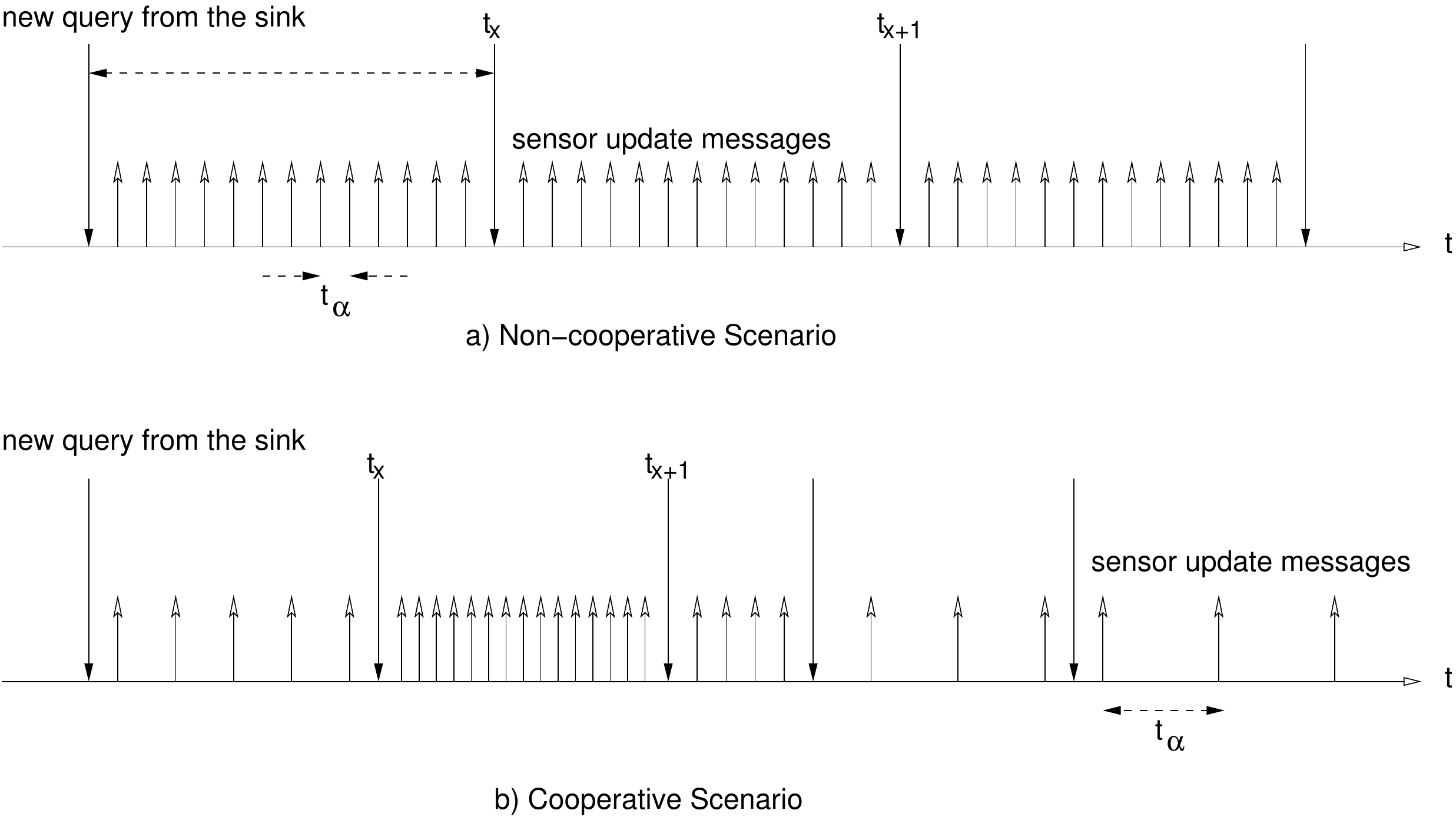}  
\caption{Non-cooperative and cooperative scenario. In the cooperative scenario, the queries are used to change the operation of the WSNs to adapt to environment based on the information received from other WSNs.}
\label{fig:Scenario}
\end{figure*}

In this section we describe how we aim to improve energy consumption and sensing quality by sharing information between WSNs. To this end, we develop a basic model, and then apply it to both of these performance indices.

We consider a set of sensor networks, consisting of WSN A ($S^A$) up to WSN N ($S^N$), with each of these networks consisting of a different number of sensor nodes. If we define every individual sensor network $S^I$ as its set of sensors $S^I = \left\{S^I_1, S^I_2, S^I_3, S^I_4, ....S^I_{n_I}\right\}$, which has cardinality $n_I$, we can express those individual sensor network sets as 

\begin{eqnarray*}
S^A & = & \left\{S^A_{1}, S^A_{2}, S^A_{3}, S^A_{4},..... S^A_{n_a}\right\} \\ 
S^B & = & \left\{S^B_{1}, S^B_{2}, S^B_{3}, S^B_{4},..... S^B_{n_b}\right\} \\
& \hdots & \\
S^N & = & \left\{S^N_{1}, S^N_{2}, S^N_{3}, S^N_{4},..... S^N_{n_N}\right\} \\
\end{eqnarray*}

We will now first consider the case of WSN $S^I$ in a standard, non-cooperative case, i.e., $S^I$ operates autonomously without any additional information. We model the operation of this network in terms of the messages sent and received by the sink of $S^I$. We consider that the sink updates the configuration of the sensors $S^I_k$ by sending query messages at time instances $t_k$, cf. Figure \ref{fig:Scenario}.a). Among other things, these queries configure the time interval $t_\alpha$ between two consecutive reports sent by the sensors to the sink. As well, it configures the number of active nodes $|S^{I}_{active}|$ in $S^I$, i.e., the subset of $S^I$ which will report its values.

We can assume $t_\alpha$ and $|S^{I}_{active}|$ to remain constant between two consecutive sink queries at $t_x$ and $t_{x+1}$. Depending on the WSN, this value may even never be changed over the lifetime of the network. In any case, $S^I$ can in the best case only take into account its own measured information to modify $t_\alpha$ and $|S^{I}_{active}|$ over time.

In our cooperative scenario, however, we assume that more information is available, coming from other WSNs, e.g., $S^J$. This additional information will allow to judge better the necessity for frequent reports and the number of active sensors. For example, if $S^J$ reports an incident, $S^I$ can decrease $t_\alpha$. In contrast, if $S^J$ reports uncritical values, $S^I$ may decrease its level of operation as well by increasing $t_\alpha$ and only requesting reports from a lower number $|S^{I}_{active}|$ of its sensors.

Thus, the situation shown in Figure \ref{fig:Scenario}.a) changes. $S^I$ may now additionally update the reporting interval every time it receives information from $S^J$, cf. Figure \ref{fig:Scenario}.b). Using a decision making process described in the next section, it can take into account its own measured information plus the information from $S^J$. If this process returns a changed $t_\alpha$, the sink of $S^I$ will issue an additional query to its sensors.

In general, we imagine that sensor nodes from both $S^I$ and $S^J$ will sense information and forward them towards their respective sinks. The sinks will then forward this information towards interested recipient WSNs, using the architecture described in the next section. However, if no information from other networks is received, the individual WSNs can still operate as in the non-cooperative scenario.

The information exchange will be useful when the exchanging WSNs sense the same parameter in the same area. However, the more interesting case is when $S^I$ and $S^J$ sense different parameters that are tightly correlated in nature, e.g., the traffic and pollution levels in the same city. Then, information about one parameter, e.g., traffic, can be used to predict the level of the other, e.g., pollution, to a certain degree.
 
Our basic assumption is that a sensor network is not interested in measuring all levels of its sensed parameter with equal interest. Instead, extreme or critical values have a much higher impact than 'normal' levels, and are therefore of more interest. As an example, high levels of pollution should be sensed with a higher accuracy than low levels or a level of 0, due to their health implications. Similarly, it is more critical to have more exact information about the location and spread of a forest fire when it occurs, than measuring with a high level of accuracy that there is no forest fire. As we will describe in the following, being able to predict the level of the sensed value should allow for a better resource usage in the affected WSN.

\subsection{Energy Consumption}

Since every sent and routed information update containing measured values consumes an amount of energy of the sensor nodes, we see that the energy consumption of the sensor network depends on $t_\alpha$ and $|S^{I}_{active}|$. For longer values of $t_\alpha$, less energy is consumed over a longer timespan, for shorter values of $t_\alpha$, more energy is consumed. Similarly, if the sink requests that a lower number of sensors $|S^{I}_{active}|$ in its network reports values, less energy will be consumed within the complete network, extending its lifetime.

Therefore, if $S^I$ receives information from $S^J$ and decides, based on this information, that it can reduce its level of operation (increase $t_\alpha$ and decrease $|S^{I}_{active}|$), it can reduce its energy consumption during that period. When a higher level of operation is necessary afterwards, e.g., because $S^J$ or its own sensors report an important event, $S^I$ can increase $|S^{I}_{active}|$ and decrease $t_\alpha$ again. Thereby, $S^I$ can adapt its energy consumption better to the current situation in the real world.

\begin{figure*}[!th]
\centering
\includegraphics[width=0.7\textwidth]{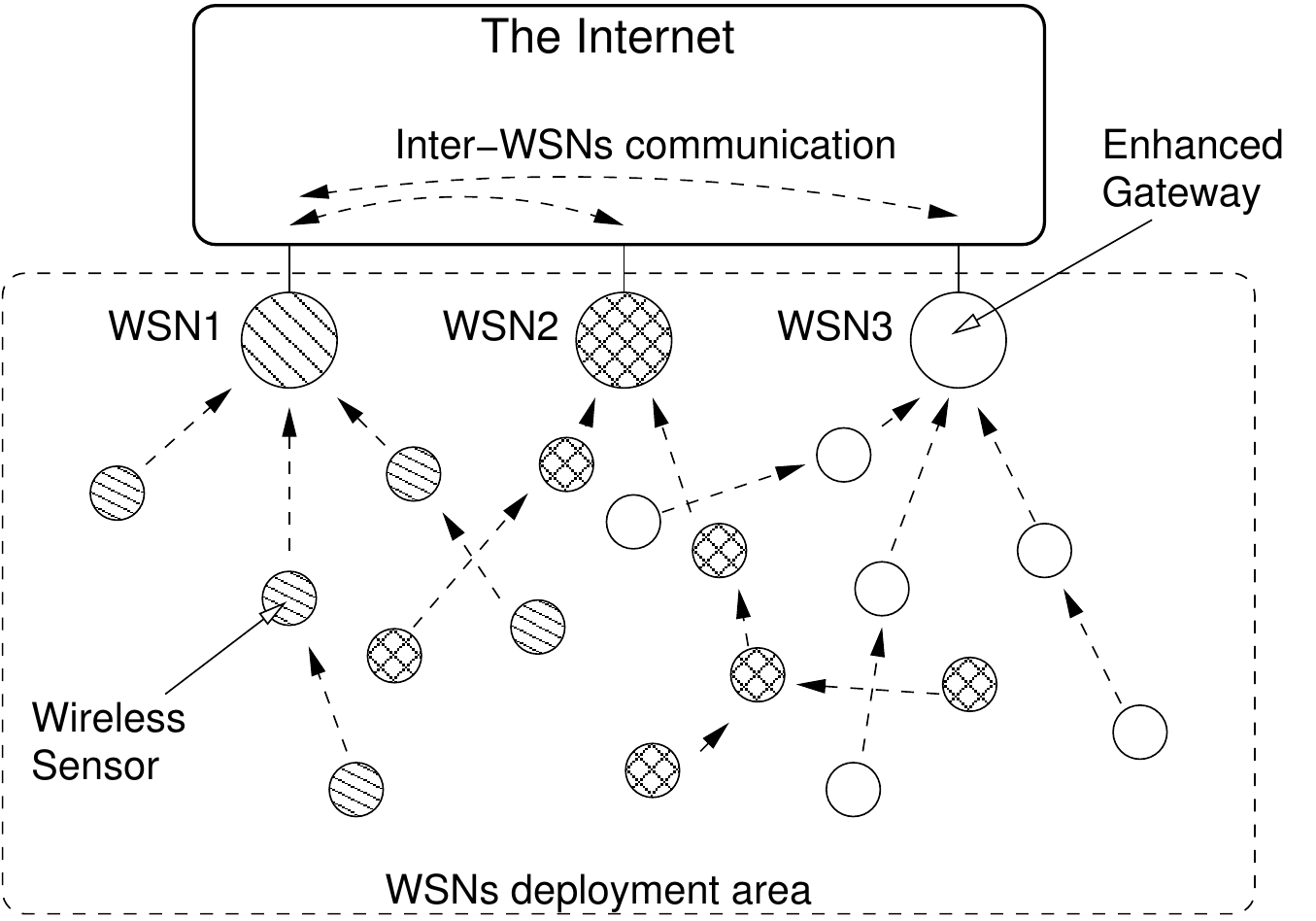}  
\caption{System Architecture}
\label{fig:System_Architecture}
\end{figure*}

\subsection{Sensing Quality}

Similarly, $S^I$ can adapt at the same time its sensing quality ($Q$). Sensing quality can be defined as the relation between the actual update frequency of sensors $\lambda = \frac{1}{t_\alpha}$ and their number $|S^{I}_{active}|$ on the one hand, and the ideal frequency $\lambda^{\ast}$ and number $|S^{I}_{active}|^{\ast}$ on the other. A simple approach for the sensing quality would therefore be

\begin{equation}\label{Eq:Q}
Q = min \left(1, \frac{\lambda}{\lambda^{\ast}} \frac{|S^{I}_{active}|}{|S^{I}_{active}|^{\ast}}\right)
\end{equation}
where the maximum achievable quality is $Q=1$. 

The values of $\lambda^{\ast}$ and $|S^{I}_{active}|^{\ast}$ depend on the current circumstances in the real world. For example, under critical circumstances (extremely high levels of traffic or temperature, an earthquake or forest fire happening), these values can be expected to be higher than under non-critical circumstances. Generally, the more sensor nodes send their sensed information to their sink, and the higher the frequency of these updates is, up to the ideal values, the better will be the sensing quality. However, this has to be paid for in terms of energy consumption. In our approach, we want therefore to improve on the utilization of this trade-off utilizing the information from additional networks.

Note that from Equation \ref{Eq:Q}, it is clear that $\lambda$ and $|S^{I}_{active}|$ should be always equal to $\lambda^{\ast}$ and $|S^{I}_{active}|^{\ast}$ to get $Q=1$. Then, if $\lambda^{\ast}$ and $|S^{I}_{active}|^{\ast}$ change with time, $\lambda$ and $|S^{I}_{active}|$ should also be adapted to those changes. Higher values than the optimal only result in a higher energy consumption, but not in a higher quality.


\section{Architecture Description}\label{sec:arch}

In this section, we will describe the components and algorithms of our architecture. These additional building blocks enable the information exchange between WSNs and the usage of this information to improve the performance of individual WSNs.

In order to interconnect multiple sensor networks, we are introducing a new entity called \textit{Enhanced Gateways} (EG) in each sensor network, as shown in Figure \ref{fig:System_Architecture}. Each EG has a direct connectivity to the sink of its sensor network, and additionally is connected to the Internet.

In some previous works this kind of entity has been shown only as Gateways, with the sole purpose of providing a bridge between the sensor networks and the IP network. In our scenario, this entity offers more than gateway functionality, as we will describe in the following. Since we are placing EGs on the edge of the fixed-line part of the Internet, we assume they do not have any energy or performance constraints.

\begin{figure*}[t]
\centering
\includegraphics[width=1.0\textwidth]{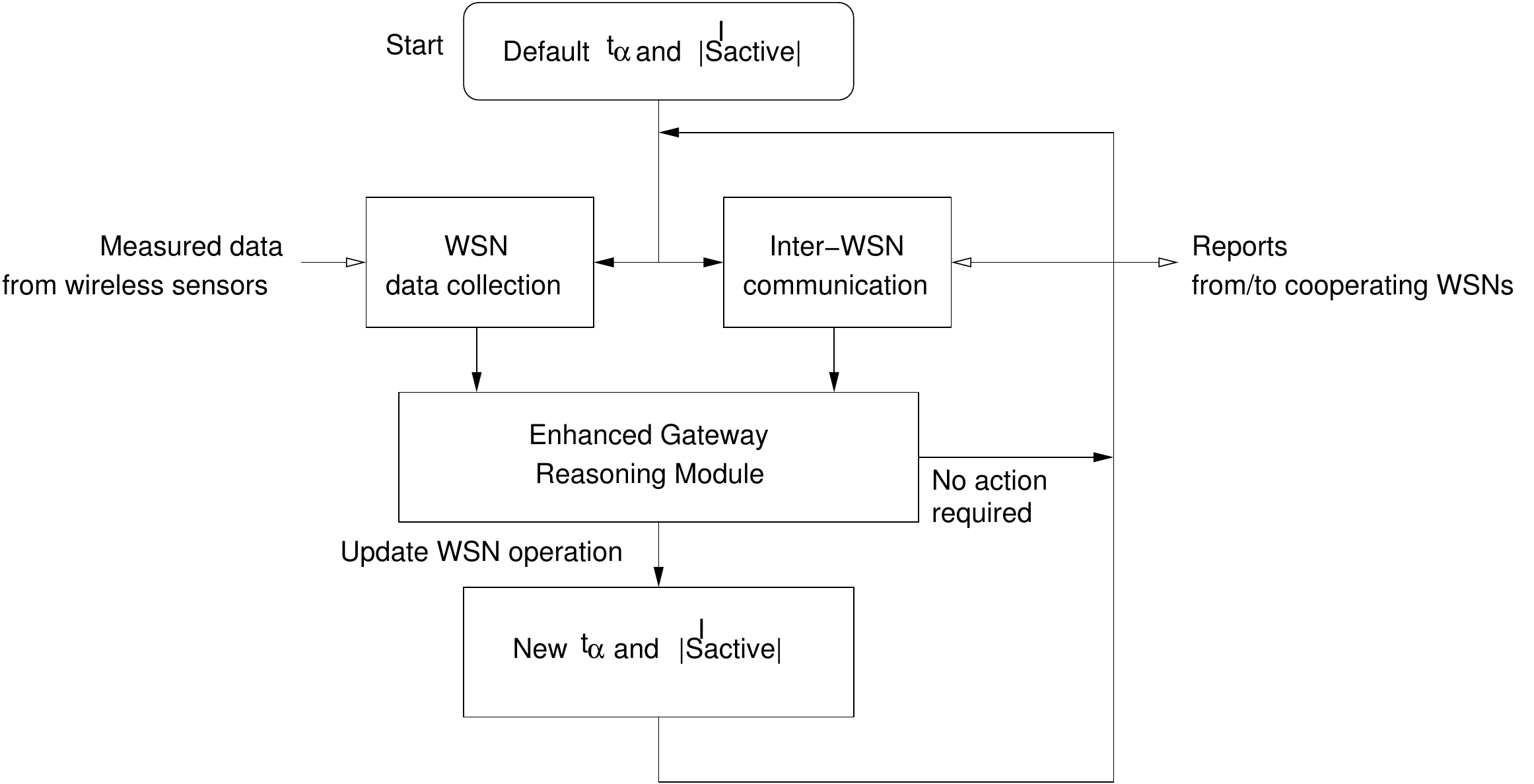} 
\caption{Enhanced Gateway Operation}
\label{fig:Flow Chart}
\end{figure*}

\subsection{Enhanced Gateway}\label{subsec:eg}

The Enhanced Gateways are the main new architectural component that our approach adds to WSNs. They are well-dimensioned machines having a reliable Internet connection, and are directly connected to their respective sensor network sink, i.e., there is one EG per sensor network. The set of all EGs maintains an overlay between themselves, as described in Section \ref{subsec:overlay}. This overlay is used to establish communication between pairs of EGs, of which at least one can benefit from information exchange.

These pairs of EGs will exchange aggregated information from their attached sensor networks in the form of regular updates. Currently, we envision this to be an average of the sensed values, but it might also be the complete set of measurements from all sensors in the network. The receipt of such an update by an EG leads to an evaluation of all available information, and possibly a reconfiguration of the attached sensor network, as described in Section \ref{sec:benefits}. Moreover, in some specific scenarios these updates may also trigger a forwarding towards other concerned entities, e.g., emergency services. 

In our architecture, we propose to evaluate received information based on the flow chart shown in Figure \ref{fig:Flow Chart}. Whenever a new update from a remote sensor network or the local sensors is received, the EG Reasoning Module considers this new information along with the latest values stored for all other information sources. Based on the trust value of all these sources, cf. Section \ref{subsec:trust_level}, and on the currency of their information, it will calculate a new level of operation, i.e., $t_\alpha$ and $S^I_{active}$. If the values of these parameters have changed in comparison to the current sensor network configuration, the EG will instruct its attached sink to generate a new query with the updated parameters and send it to the sensors.

\begin{table*}[!t]
\centering 
\caption{Operation Level of a WSNs measuring air pollution when the measured road traffic is used to trigger it.}
\begin{tabular*}{1.0\textwidth}{@{\extracolsep{\fill}}lcp{2.5cm} p{3cm}}
\hline
Condition & Operation Level \\
\hline
\texttt{If $P \leq 10 \mu g/m^3$ \&\& $T \leq 10$ veh./min} & Pollution Network Operation Level = Low \\
\texttt{If $10 \mu g/m^3 < P \leq 20 \mu g/m^3$ \&\& $10$ veh./min $< T \leq 50$ veh./min} & Pollution Network Operation Level = Moderate \\
\texttt{If $10 \mu g/m^3 < P \leq 45 \mu g/m^3$ \&\& $10$ veh./min $< T \leq 50$ veh./min} & Pollution Network Operation Level = High\\
\hline
\end{tabular*}
\label{tab:rules}
\end{table*}

The calculation of the level of operation should, e.g., take into account if the local sensor network measured highly variable or extreme values, and whether reported values for correlated parameters indicate a change in the measured value soon. To illustrate this, we will give in Table \ref{tab:rules} some example rules that we will use in a straightforward rule-based approach for this algorithm for measured values of pollution ($P$) and traffic ($T$).

\begin{figure*}[!t]
\centering
\includegraphics[width=0.7\textwidth]{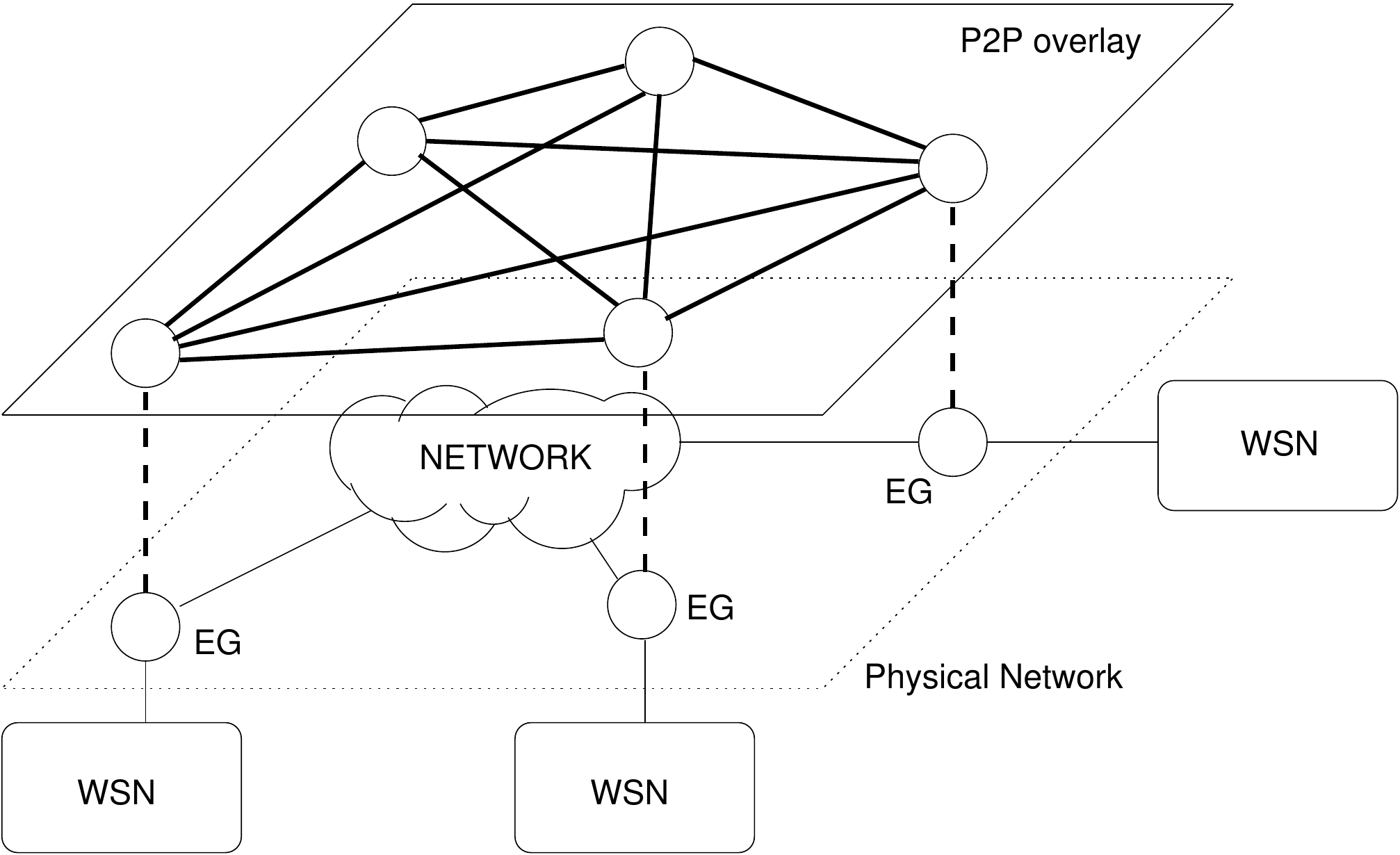} 
\caption{P2P overlay Network}
\label{fig:P2Poverlay}
\end{figure*}

\subsection{Overlay}\label{subsec:overlay}

The EGs $E^I$ of all participating sensor networks $S^I$ will be maintaining a single-hop P2P overlay between themselves as shown in Figure \ref{fig:P2Poverlay}. The primary reason for selecting a single hop overlay is that we want to keep the lookup time to locate other corresponding EGs as low as possible. As in our scenario the nodes are stable and not changing their locations, the churn rate will be low, making the use of such an overlay feasible.

In a one-hop DHT, every EG will hold some state about every other EG. This state is contained in a \textit{Global Lookup Table} (GLT), which will hold a small number of values for each EG and their sensor network. These values are the overlay node identifier, the IP address of the corresponding enhanced gateway, the GPS coordinates of the centre of the sensor network, and finally the network category, identifying what kind of information the sensor network is collecting. Thus, the GLT looks like shown in Table \ref{tab:glt}. 

An EG wishing to join the overlay generates an attach request message and sends it to an already participating node, which will respond with its GLT. The address of a participating node can be obtained using a globally available information source, such as a web service. As well, this step may include a registration or authentication process.

\begin{table*}[!t]
\centering 
\caption{Global Lookup Table}
\begin{tabular*}{1.0\textwidth}{@{\extracolsep{\fill}}lccp{2.5cm} p{3cm}}
\hline
Nodes & IP Address & Node ID & Co-ordinate & Network Category \\
\hline
$EG^A$	& 132.187.16.21      & CF32A1            & N 49$\,^{\circ}$ 47' 39.4506", E 9$\,^{\circ}$ 55' 38.9778"                 & Pollution \\
$EG^B$	& 141.101.126.147      & 63D80B            & N 41$\,^{\circ}$ 23' 16.6416", E 2$\,^{\circ}$ 10' 11.715"                 & Humidity \\
$EG^C$  & 193.174.81.220     & 4248C4            & N 49$\,^{\circ}$ 47' 39.4510", E 9$\,^{\circ}$ 55' 38.9703"                  & Traffic \\
...     & ...      & ....            & ....                 & .... \\
...		& ...      & ....            & ....                 & .... \\
...		& ...      & ....            & ....                 & .... \\ 
...		& ...      & ....            & ....                 & .... \\
$EG^N$  & 195.251.255.138      & 815162            & N 37$\,^{\circ}$ 58' 44.8932", E 23$\,^{\circ}$ 42' 59.688"                 & Pollution \\
\hline
\end{tabular*}
\label{tab:glt}\end{table*}

\subsection{Cooperating Networks}

While the GLT means that an EG knows about the existence of all other sensor networks willing to exchange information, it is neither feasible nor necessary to actually establish a cooperation with all of them. Only sensor networks that are roughly in the same area in the physical space, and that gather information about related parameters are of interest for communication. Therefore, a second list of EGs needs to be derived from the GLT, containing the sensor networks with which the local EG wants to share information. We will call this secondary list Cooperating Networks Table (CNT).

The formation of this secondary table is based on the aim that it should consist of only those networks who will qualify for cooperation. As a simple algorithm to implement this functionality, we propose to use the physical proximity between sensor networks, as discussed in Section \ref{subsubsec:network_distance}. Apart from this distance calculation, every EG also should make its selection based on the network category column of the GLT, which allows to filter for sensor networks measuring a related parameter.

While this simple algorithm allows to rule out communication with unsuited networks, not all sensor networks included in the CNT will provide the same support in terms of useful information. Thus, we propose to assign a trust level to each of these networks, reflecting the value of the provided information to the local performance optimization. A first approximation of this trust can be made based upon the distances between the networks, as we describe in Section \ref{subsec:trust_level}. In Section \ref{subsubsec:coeff}, we will describe another, experience-based method to fine-tune the trust level. 

This value, along with the currency of received information, should be taken into account when making decisions based on information coming from other networks, cf. Section \ref{subsec:eg}. Thus, the CNT has the form seen in Table \ref{tab:cnt}. The time interval signifies how frequently updates are received from the other sensor network, while latest value holds the last reported information. Finally, the CNT holds the timestamp at which the last update containing that information arrived. 

\begin{table*}[!t]
\centering 
\caption{Cooperating Networks Table}
\begin{tabular}{lcccc}
\hline
Nodes & Trust Value & Time Interval  & Latest Value & Timestamp \\
\hline
$EG^A$	& 6      & 600            & 35 $\frac{\mu g}{m^3}$                & 10:03h \\
$EG^C$	& 8      & 300            & 10 $\frac{1}{min}$                 & 10:05h \\
...		& ...      & ....            & ....                 & .... \\ 
...		& ...      & ....            & ....                 & .... \\ 
$EG^J$  & 2      & 300            & ....                 & 10:07h \\
\hline
\end{tabular}
\label{tab:cnt}
\end{table*}

\subsubsection{Calculating Sensor Network Distances}\label{subsubsec:network_distance}

In this section, we will explain how to calculate the distance between two sensor networks, as this value is used to form the CNT.

We consider $(x_I,y_I)$ as the geographical location coordinates for sensor network $S^I$. Typically, we assume these coordinates to be the geographical center of the sensor network. It is one of our assumptions that while deploying the sensor network, the concerned person will note these location coordinates.

Since these locations are part of the GLT, the local EG forming its CNT can calculate the distance between its own sensor network center and the coordinates of all other sensor networks. This distance calculation between two GPS coordinates can be performed using the Haversine distance calculation formula shown in Equation \ref{Eq:distance_formula}, where $d$ is the distance between 2 geographical location coordinates, $R$ is the mean radius of Earth (i.e. $6371$ km), ${\bigtriangleup lat}$ is the difference between $lat2$ and $lat1$ and ${\bigtriangleup long}$ is the difference between $long2$ and $long1$. $lat1$ ($long1$) is the latitude (longitude) of the network computing $d$ and $lat2$ ($long2$) is the latitude (longitude) of the other network.

\begin{figure*}[t]
\begin{eqnarray}\label{Eq:distance_formula}
a & = & \sin^{2}\left(\frac{\bigtriangleup lat}{2}\right) + \cos({lat1})\cos({lat2})\sin ^{2}\left(\frac{\bigtriangleup long}{2}\right) \\
d & = & 2 R \arcsin{\left(\sqrt{a}\right)} \nonumber
\end{eqnarray}
\end{figure*}

Using these distances, one for every sensor network participating in the global architecture, the local EG can now filter out all networks that are farther away than a threshold $d^I_{max}$. This threshold can depend on the type of the local sensor network and is therefore not a globally fixed value. Only networks closer than this maximum distance are considered for inclusion in the CNT. In addition, their category must match, as described in the following section.

\subsubsection{Matching Network Category}

After calculating the distance, the EG evaluates the network category of the remaining networks. This value will show what parameters those networks are sensing. Every EG should know with what other kind of network it can collaborate with, e.g., by pre-defining it in the EGs' software. If the category of the remote sensor network matches, it can be included in the CNT.

\subsection{Level of Trust}\label{subsec:trust_level}

As described, with the trust value we aim to give an EG the option to differentiate between information sources, and to weigh their input. Any EG that is part of the CNT has a base trust value, since information from this entity is accepted. However, EGs should be trusted more the more is known about them and the quality of their information.

As a first default value for the trust of other EGs, without any more data, we can use the proximity of their attached sensor network center to the local sensor network coordinates. This allows us to linearly scale the trust value between a maximum initial value, e.g., 10, for having the exact same coordinates, and a value of 0 for a distance of $d^I_{max}$.

If, additionally, the coordinates of the individual sensors of the two networks are available as well, a more fine-grained distance could be computed, with an according higher trust value if this calculation shows a good match between the two networks.

Finally, we can increase the trust value of remote networks by keeping track of their reports and how well they correlate with the measured values in the local sensor network, as described in the following.

\subsubsection{Correlation Coefficient of Measured Information}\label{subsubsec:coeff}

\begin{figure*}
\begin{equation} \label{Eq:correlation}
r_{v^I v^J} = \frac{n\sum v^I v^J - (\sum v^I)(\sum v^J)}{(\sqrt{n\sum v^{I^2}- (\sum v^I )^{2}})\cdot(\sqrt{n\sum v^{J^2}-(\sum v^J)^{2}})} 
\end{equation}
\end{figure*}

As described, the EG of a sensor network $S^I$ will regularly receive reports from each cooperating EG, e.g., $S^J$. We define the contained reported value of the report number $y$ as $v^J_y$. $EG^I$ can also note the current value reported by the local sensor network, $v^I_y$, and thus create a pair of values for each received report. Over time, it thus can create a combined history of externally and internally collected information, i.e., $v^I$ and $v^J$.

Once enough values are collected, $EG^I$ can calculate the correlation coefficient function $r_{v^I v^J}$ using Equation \ref{Eq:correlation}, where $n$ is the length of the recorded history.

The trust value can then be increased if a strong positive or negative correlation is seen, i.e., by common definition for $|r_{v^I v^J}| > 0.7$. Whether we see a positive or negative correlation depends on the scenario, i.e., traffic levels and pollution should generally be positively correlated, while humidity and forest fire probability should be negatively correlated.

In this experience-based approach, $r_{v^I v^J}$ can be updated over time. Thus, an EG can react to changes in sensor deployments, better coverage, etc. Remote EGs that report more and more useful information will thus be able to 'earn' more trust, and their reports will then have a higher weight in the local decision making process.


\section{Use Case Descriptions}\label{sec:usecases}

In this section, we will briefly describe two scenarios where we believe our approach could be beneficial. However, these examples are merely illustrative and not exhaustive, since our approach works for any combination of sensed information that is correlated in nature, such as humidity and rainfall, temperature and agricultural growth monitoring, or earthquakes and structural monitoring.

\subsection{Traffic \& Pollution}

\begin{figure*}[t]
\centering
\includegraphics[width=1.0\textwidth]{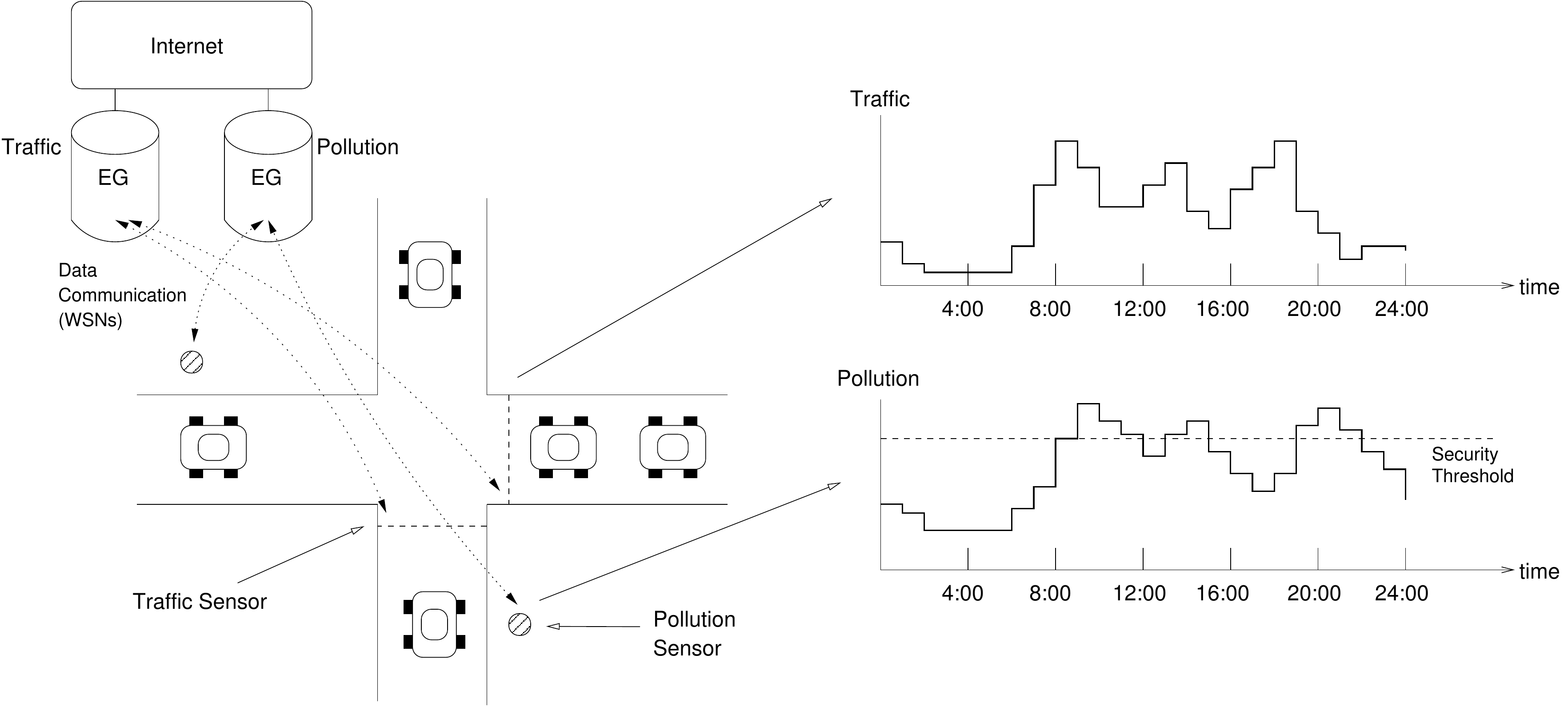} 
\caption{Traffic \& Pollution Scenario}
\label{fig:Figure5}
\end{figure*}

Traffic and pollution are two basic areas where we can find wide spread usage of sensor networks in metropolitan cities around the world. Traffic Sensors are normally widely deployed for traffic management and monitoring by the civil authorities. These sensors can measure vehicle quantities passing by during a fixed period of time, and in addition they may monitor vehicle classes and velocity. This scenario is depicted in Figure \ref{fig:Figure5}.

Pollution sensors, on the other hand, are used especially in large cities to keep track of the smog generated there by the high concentration of industry and traffic. Since especially the day-to-day air pollution has health implications, data from these sensor networks can be used to warn citizens with a form of pollution scale, or traffic may be managed differently to reduce the amount of cars in the city. 

Since a significant part of the air pollution in large cities is caused by traffic, we envision that a cooperation between these types of sensor networks could be beneficial. Since the level of traffic typically does not depend on the level of pollution, there will be mainly an information flow from the traffic sensor network EGs to the pollution sensor network EGs. Each update contains the currently measured traffic levels. The EG of the receiving pollution sensor network will therefore receive information about rising traffic levels, and can then increase the level of operations of its network accordingly to better sense an upcoming peak in pollution. On the other hand, if low levels of traffic are reported, and if the pollution sensor themselves show no critical values, the EG might decide to increase $t_\alpha$ and decrease $|S^I_{active}|$.

\subsection{Temperature, Wind speed, Humidity \& Forest Fire Probability}

\begin{figure*}[t]
\centering
\includegraphics[width=1.0\textwidth]{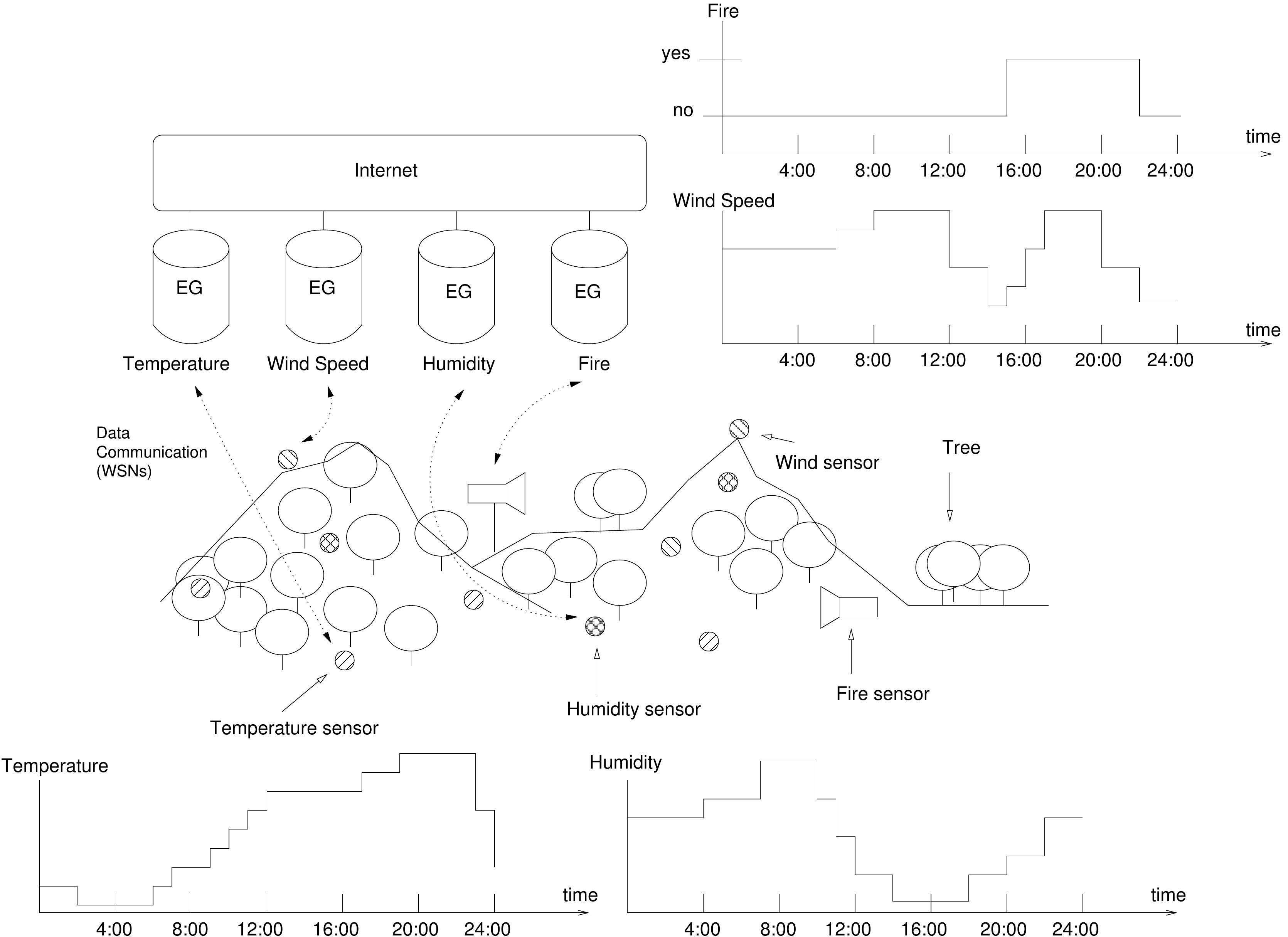} 
\caption{Temperature, Wind speed, Humidity \& Forest Fire Probability}
\label{fig:Figure6}
\end{figure*}

The previous scenario illustrates an example for one type of sensor network using data of a second type. However, our architecture allows a more general cooperation, i.e., the collection and utilization of data from a number of sources of different kinds. An example for this is a forest fire detection sensor network, as depicted in Figure \ref{fig:Figure6}.

Geographical information datasets have shown that the probability for forest fires is directly proportional to specific environmental conditions, such as very high temperature, extremely low relative humidity and very high wind-speed \cite{ref-journal15}. As an example we can take the Black Saturday Bush Fire of 2009 in Australia. According to records, this incident was made possible by the presence of all of the above mentioned favorable conditions. Temperatures reached $46.4$ degrees Celsius, while humidity levels as low as $6$\% were reported, and wind speeds of $100$ km/h were measured \cite{ref-journal14}. All three of these metrics can be and are monitored by sensor networks, e.g., for meteorological purposes.

Thus, we believe the information gathered by these types of sensor networks can be used to adapt the level of operation of a forest fire detection sensor network. In periods with low temperatures and high humidity, the level of operation of the detection network can be decreased because the probability for a fire is very low. On the other hand, if high temperatures and a low humidity are reported in the general area for a longer period, and if in addition high wind speeds are measured, then the EG should let more sensors report in shorter intervals to be able to detect a fire quickly in this critical situation.


\section{Other Initiatives}\label{sec:relwork}

As we have mentioned, traditionally research in WSNs has been focused on internal issues like routing, self organization, new MAC or routing protocol development, and energy consumption. Due to the numerous approaches in these fields, WSNs show a heterogeneous usage of protocols and mechanisms. Integrating these heterogeneous networks with the IP world has become a new research direction recently. In this field, some work has been done regarding architectures for interconnecting WSNs. 

To interconnect multiple spatially distributed heterogeneous sensor networks, two basic approaches exist, namely a client/server-based approach and a P2P approach. As in other application fields, both approaches have their advantages and disadvantages, i.e., the more centralized system offers a higher level of control, but at the price of introducing a central point of failure and scalability issues. Due to the fact that our approach belongs to the class of P2P architectures, we will primarily discuss these systems.

IrisNet \cite{ref-journal2} was one of the first large scale overlay-based architectures. It was primarily meant to interconnect various multimedia sensors which are distributed world-wide and which are capable of running rich application suites. Sharing infrastructure resources is the most important design principle of the IrisNet architecture. 

Hourglass \cite{ref-journal3} is an Internet-based software architecture to connect various sensor networks, their services and applications. Hourglass consists of an overlay network of dedicated connected machines which provides service registration, node discovery and routing of data from sensors to clients application. 

SharedSense \cite{ref-journal4} is mainly a peer to peer environment for connecting multiple heterogeneous sensor networks and monitoring them. It is based upon JXTA \cite{ref-journal5}, which is a Java based P2P substrate, having a different set of rules and regulations in comparison to other popular P2P overlays \cite{ref-journal6}. 

The authors of \cite{ref-journal7} proposed a mobile P2P sensor networks overlay which was based upon 3G mobile networks. This overlay was also based upon JXTA. P2PBridge \cite{ref-journal8, ref-journal9} was another initiative with the idea to have an inter WSN-communication and interoperability. Again, JXTA is used as the P2P substrate. 

The MetroSense Project \cite{ref-journal10} is a general purpose heterogeneous architecture based on people-centric sensing applications. GSN \cite{ref-journal11}, or the Global Sensor Network platform is a flexible middleware approach which was envisioned to provide a common platform to connect multiple widely deployed sensor networks and create a global Sensor Internet. GSWSN \cite{ref-journal12}, or Global Scale Wireless Sensor Networks is another Internet-based overlay architecture to interconnect globally dispersed heterogeneous wireless sensor networks. Finally, in \cite{ref-journal13}, a hierarchically structured worldwide sensor web architecture has been proposed to retrieve data from multiple heterogeneous sensor networks.

All these architectures, and comparable server-based approaches, have in common that they interconnect different and heterogeneous sensor networks. However, the functionality of these architectures mainly consists in enabling a global management and data evaluation. Information is generally envisioned to flow from sensor networks towards users or external systems. In contrast, our architecture has the primary goal of enabling communication and cooperation between the sensor networks themselves, with the specific goal of performance optimization for individual WSNs.


\section{Conclusion and Open Research Directions}\label{sec:conc}

In this paper, we have proposed a new concept to improve the performance of WSNs based on information received from other sensor networks, opening new research directions in both the WSN and Networking community. We described the internal WSN mechanisms used for this optimization and our reasoning for their beneficial effect. In addition, we presented an architecture and algorithms enabling this information exchange and the processing of received data. 

To further develop the proposed concept, there are several topics which require special attention:

\begin{itemize}
	\item Create a complete list with the WSNs combinations (scenarios) that can benefit from our proposed architecture. The details for each combined scenario have to include the types of sensor networks that will be sharing information between themselves, the criteria to be met for any kind of information sharing to be practically feasible, the different operation levels in every individual sensor network, the kind of communication messages to be exchanged, etc.	
	\item Develop new reasoning models based on the correlation of the sensed values from each individual WSN collaborating. This relationship model should be constructed based upon real-life geographical data. It has to provide the understanding about how the measured values of those environmental parameters do actually relate to each other, and therefore form the basis for the decision-making process in the EGs. 	
	\item Evaluate the functioning and the performance of this kind of collaborative architecture, as well as the protocols that compose it. Through our description in Section \ref{sec:benefits}, we have shown how can we achieve our target of performance optimization of WSNs by controlling their energy consumption and level of sensing quality, but quantitative works are required. 
\end{itemize}


\section*{Acknowledgements}

This work has been partially supported by the Spanish Government under projects TEC2008-06055 (Plan Nacional I+D), CSD2008-00010 (Consolider-Ingenio Program) and by the Catalan Government (SGR2009\#00617).


\end{document}